\documentclass[prb,twocolumn, superscriptaddress,showpacs,floats]{revtex4}

\usepackage[latin1]{inputenc}
\usepackage[intlimits,tbtags]{amsmath}
\usepackage{color}
\usepackage{amsfonts}
\usepackage{amssymb}
\usepackage{graphicx}
\usepackage{bm}

\newcommand{\C}[2]{\ensuremath{C_{ #1}^{#2}}}
\newcommand{\Ha}{{\ensuremath{\text{Ha}^*}}}
\newcommand{\exb}{E_b^{\mbox{\tiny X}}}

\begin{document}

\title{Semiconductor quantum tubes: dielectric modulation and excitonic response}
\author{David Kammerlander}\email{david.kammerlander@unimore.it}
\affiliation{CNR-INFM Research Center for nanoStructures and
bioSystems at Surfaces (S3)}
\affiliation{Dipartimento di Fisica,
Universit\`a di Modena e Reggio Emilia, Via Campi 213/A, 41100
Modena, Italy}
\author{Filippo Troiani}
\affiliation{CNR-INFM Research Center for nanoStructures and
bioSystems at Surfaces (S3)}
\author{Guido Goldoni}
\affiliation{CNR-INFM Research Center for nanoStructures and
bioSystems at Surfaces (S3)}
\affiliation{Dipartimento di Fisica,
Universit\`a di Modena e Reggio Emilia, Via Campi 213/A, 41100
Modena, Italy}

\date{\today}

\begin{abstract}
We study theoretically the optical properties of quantum tubes, one-dimensional semiconductor nanostructures where electrons and holes are confined to a cylindrical shell. In these structures, which bridge between 2D and 1D systems, the electron-hole interaction may be modulated by a dielectric substance outside the quantum tube and possibly inside its core. We use the exact Green's function for the appropriate dielectric configuration and exact diagonalization of the electron-hole interaction within an effective mass description to predict the evolution of the exciton binding energy and oscillator strength. Contrary to the homogeneous case, in dielectrically modulated tubes the exciton binding is a function of the tube diameter and can be tuned to a large extent by structure design and proper choice of the dielectric media.
\end{abstract}



\maketitle

\section{Introduction}

Cylindrical semiconductor nanostructures bridge between quasi-1D systems at small diameters and quasi-2D in the opposite limit, thus extending the wealth of physics
and applications of low-dimensional solid-state systems.
The controlled growth of semiconductor quantum tubes (QTs) with diameters in the
10-100 nm range has been recently demonstrated through several techniques,
including multi-layer overgrowth of nanowires\cite{lauhon:02,mohan:06,fontcuberta:08}
and strain-induced bending of a planar heterostructure.\cite{prinz:00,schmidt:01}
In addition to QTs with a solid semiconductor core, it is possible to grow \emph{hollow} QTs, where the charge carriers are confined in a thin semiconductor shell, encompassed
by a barrier material which is only a few nm thick.\cite{noborisaka:05, mohan:06b, golod:01} Large surface-to-volume ratios and the possibility of various
functionalizations on both the internal and external surfaces make the latter systems particularly interesting for applications.\cite{lieber:03}

Although experiments concerning the optical properties of these systems are still
limited, advancements in the optical quality of the samples point to a rapid increase
of these investigations.\cite{mohan:06, pal:08, goto:09, jabeen:08} The excitonic properties of semiconductor QTs are particularly interesting with respect to conventional semiconductor quantum wires, where excitons are confined in the core of the
nanostructure.\cite{nagamune:92, wegscheider:93, someya:95, rossi:96, rossi:97, katz:02, duan:03, slachmuylders:06} On the one hand, due to the combined effect of the QT curvature and
of the quasi-2D confinement of carriers in the cylindrical shell, excitonic binding
energies might be substantially stronger than in bulk, even for large diameter QTs.
On the other hand, a dielectric medium outside the shell of the QT may result in a dielectric confinement of the electric field felt by the optically excited electron-hole pairs, in most cases enhancing their excitonic binding energy. Since the dielectric interface is spatially separated from the carriers, which are confined deep inside the shell, excitonic binding and sensitivity to the medium might be strongly enhanced without spoiling the optical properties of the electronic system,\cite{goldoni:98} analogously to core-shell nanowires.\cite{jabeen:08}
The screening provided by the dielectric environment can be varied in a broad range.\cite{issac:05} The tunability of the dielectric constant in the core of the QT, obtained,
{\it e.g.}, by oxidation,\cite{fiore:98} can further increase such effects.

Present work on QTs theoretically considered magnetic states,\cite{ferrari:08, ferrari:09} reported experimental evidence of the Aharonov--Bohm effect,\cite{nomura:08} and treated optical properties,\cite{nomura:07} but the influence  of the dielectric dismatch between the nanostructure and the environment has been studied so far only for conventional quantum wires\cite{goldoni:98} and freestanding nanowires.\cite{slachmuylders:06} Here we will consider also a dielectric mismatch between the core and the shell, which will lead to a considerable change in the electron-hole interaction, as shown in Fig.~\ref{fig:potential}. 

Hereafter we investigate the excitonic binding and oscillator strength in hollow and
filled QTs for different geometries and dielectric configurations.
Besides increasing due to the reduced screening, the excitonic binding strongly
depends on the QT diameter and on the dielectric medium.
The paper is organized as follows. In Section~\ref{sec:model} we outline the theoretical model,  which includes
the exact solution of the Poisson equation and the diagonalization of the electron-hole Hamiltonian within the
envelope-function approximation. In Sections~III and IV we report our results and draw
the conclusions, respectively.

\section{The Model}\label{sec:model}

\begin{figure}[h]
\includegraphics[width=0.6\columnwidth,clip, angle=-90]{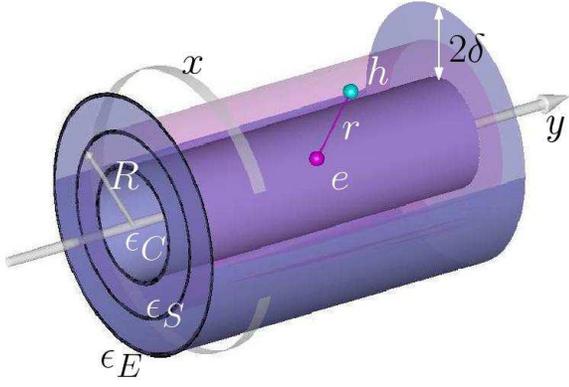}
\caption{(color online) Schematics of a QT with outer surface cut open. The electron-hole pair are constrained to a cylindrical surface of radius $R$, embedded in a shell of thickness $2\delta$. $x$ and $y$ are the relative electron-hole coordinates. The dielectric constants $\epsilon_E, \epsilon_S, \epsilon_C$ characterize the environment, the shell and the core region, respectively.}
\label{fig:tube_scheme}
\end{figure}

The system we consider consists of an infinite tube with cylindrical symmetry\cite{nota:spigoli} (see
Fig.~\ref{fig:tube_scheme}). For simplicity we assume that the motion in the radial direction is frozen, and that charge carriers are radially confined in a $ \delta $-like well at a distance $R$
from the tube axis. This electronic layer is buried in the middle of a coaxial cylindrical shell of thickness $2\delta$ with dielectric constant $ \epsilon_S $, while the core and the environmet have in general different dielectric constants, $ \epsilon_C$ and $ \epsilon_E $, respectively. Since the shell is a semiconductor material, typically $ \epsilon_S \ge \epsilon_C, \epsilon_E $.\cite{issac:05}

The invariance under translations along, and rotations around the tube axis warrants
the separation of the center of mass and relative coordinates. The motion of the Wannier exciton\cite{sham:66} in the relative degrees of freedom is determined by the envelope-function  Hamiltonian
\begin{equation}\label{eq:hamiltonian}
H (x,y) = -\frac{1}{2} \left[ \frac{\partial^2}{\partial x^2}+  \frac{\partial^2}{\partial y^2}\right] - V (x,y),
\end{equation}
expressed in units of the effective Hartree \Ha$=(\mu/\epsilon_S^2)
\text{Ha}$, with $ \mu = m_e m_h / (m_e + m_h) $ the reduced electron-hole mass. The relative coordinates around the circumference ($ x=R \phi $) and along the tube axis ($y$,
see Fig. \ref{fig:tube_scheme}), are in units of the effective Bohr radius,
a$_B^*=(\epsilon_S/\mu) \, \text{a}_B$.

The effective Coulomb interaction potential $V(x,y)$ between the confined electron
and hole depends parametrically on the dielectric constants ($\epsilon_C, \epsilon_S, \epsilon_E$) and on the tube geometry through $\delta$ and $R$.
In cylindrical coordinates, the potential (scaled with \Ha) generated by a charge at $r^\prime=(\rho^\prime, \phi^\prime, z^\prime)$ reads
\begin{align}
V_{\alpha}(\bm{r},\bm{r'})&= \frac{\epsilon_S}{2 \pi^2} \sum_{m=-\infty}^{\infty} e^{\imath m (\phi-\phi')} \times \nonumber\\
&\int_0^{\infty}dk \cos[k(z-z')] g_{m,\alpha}(k, \rho, \rho'),
\end{align}
where $ \alpha = C ,S , E$ indicates whether the position of the test-charge $r=(\rho, \phi, z)$ is in the core, shell or environment region, respectively, and $g_{m,\alpha}(k, \rho, \rho')$ is the solution of the radial Poisson equation in that region (see the Appendix for further details).
The interaction $V$ in Eq.~\eqref{eq:hamiltonian} coincides with $V_S$,
with $\rho=\rho^\prime=R$. As shown in the Appendix,
\begin{eqnarray}\label{eq:g_m}
g_{m,S}(k,R,R)  & = & \frac{4 \pi}{\epsilon_S} \left[ \tilde{B}_m^< + \tilde{C}_m^<\right] \left[\tilde{B}_m^>+C_m^>\right] \times \nonumber \\
   && I_m(kR) K_m(kR),
\end{eqnarray}
where $I_m,K_m$ are the Bessel functions of the first and second kind; the coefficients $\tilde{B}_m^<, \tilde{C}_m^<, \tilde{B}_m^>, C_m^>$ are given in Eqs.~\eqref{eq:green_fu_coeff} and \eqref{eq:substit_c_b} in terms of $ I_m, K_m $ and their derivatives.

\begin{figure}[h]
\includegraphics[width=0.98\columnwidth,clip]{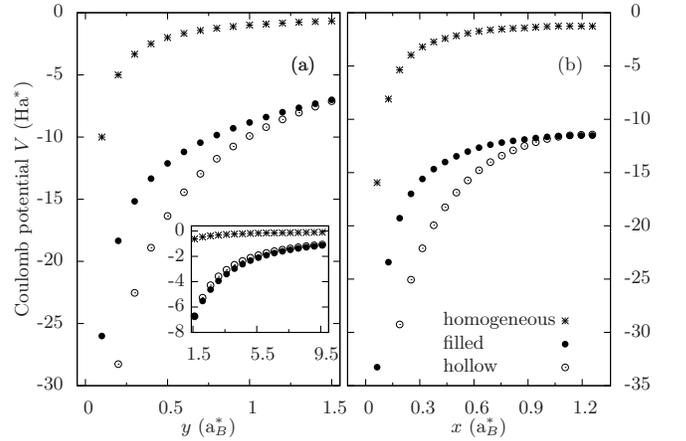}
\caption{Electrostatic interaction $ V (x,y) $ between an electron-hole pair confined to a cylindrical surface with diameter $ D=0.8 $~a$_B^* $, buried in a shell with thickness $ 2\delta=0.1 $~a$_B^*  $ with dielectric mismatch, as follows. Homogeneous case: $ \epsilon_C=\epsilon_S=\epsilon_E $. Filled case: $\epsilon_C=\epsilon_S=10 \epsilon_E$. Hollow case: $ \epsilon_S = 10 \epsilon_C = 10 \epsilon_E $. (a) interaction along the QT, $V(x=0,y)$. (b) interaction around the QT, $V(x,y=0)$. Inset in (a): $ V(x=0,y)$ in a larger range of $ y $.}
\label{fig:potential}
\end{figure}

To illustrate how the electron-hole interaction is influenced by the dielectric environment,
we shown in Fig.~\ref{fig:potential} the potential $V$ for i) a \emph{filled} QT, with a core of the same material as the shell, immersed in a substance with a low-dielectric constant ($ \epsilon_C=\epsilon_S = 10 \epsilon_E $), and ii) a \emph{hollow} QT, with the same low-dielectric constant substance inside and outside the shell ($ \epsilon_S = 10 \epsilon_E = 10 \epsilon_S$. For comparison, we also show the dielectrically homogeneous case ($ \epsilon_C=\epsilon_S= \epsilon_E $), where the $V$ reduces to the usual Coulomb potential $ V(x,y)=-1/\epsilon_S \sqrt{(2R\sin{(x/2R)})^2+y^2} $. Figure~\ref{fig:potential}(a) shows the interaction \emph{along} the QT $ V(x=0,y) $, while Fig.~\ref{fig:potential}(b) shows the interaction \emph{around} the cylinder $ V(x,y=0) $. The Coulomb interaction for the hollow and filled cases is for all distances stronger than in the homogeneous case, since the \emph{average} dielectric constant of the system is smaller, and the electric field is not screened outside and, for the hollow case, also inside the QT. The interaction in the filled and hollow case is substantially different only for distances smaller or comparable to the Bohr radius, with the interaction in the hollow case being stronger. For larger distances (inset of Fig.~\ref{fig:potential}(a)), on the other hand, the non-trivial influence of the dielectric mismatch between the core and the shell leads to crossing of the potentials for hollow and filled QTs, before both converge to the same value.

A convenient basis set to represent the exciton wavefunction is obtained by multiplying
eigenfunctions of the linear momentum operator along $y$ ($ e^{\imath k y}$) and of the angular momentum operator along the tube axis ($e^{\imath n x/R}$). Imposing periodic Born-von Karman boundary conditions\cite{born:12} along $y$, with period $L$ sufficiently larger than the effective Bohr radius of the material, results in: $k=p \Delta k$ with $ \Delta k=2\pi / L $. The wavefunction thus reads
\begin{eqnarray}\label{eq:wavefunc}
\psi_j (x, y) & = & \frac{1}{2\pi}\sqrt{\frac{\Delta k}{4R}} \times \nonumber\\
& &
\sum_{n=-N}^{N}
\sum_{p=-P}^{P}
\C{n, p}{j} e^{\imath n x/R} e^{\imath p \Delta k y},
\end{eqnarray}
where ${p,n}\in\mathbb{Z}$ and $j$ indicates the $j-$th exciton state. The coefficients $\C{n,p}{j}$ are obtained from the Schr\"odinger equation in the above basis:
\begin{multline}\label{eq:schroedinger}
\sum_{n'\!, \,p'} \bigg\{ \frac{1}{2}\bigg [ \bigg (\frac{n'}{R}\bigg)^2+
(p' \Delta k)^2\bigg] \delta_{n,n'}\delta_{p,p'}\\
-U_{{n'\!,p'\atop{\!\!n,p}}}\bigg \} \C{n'\!,\,p'}{j}=E_j \C{n,p}{j}.
\end{multline}
The diagonal term in the first line represents the kinetic energy, whereas the matrix elements of the electron-hole Coulomb interaction term are given by
\begin{equation}\label{eq:coulomb_matrix}
U_{{n'\!,p'\atop{\!\!n,p}}}= \frac{\Delta k}{(2\pi)^2} g_{|n-n'|,\,S}\,(|p-p'|\Delta k,R,R).
\end{equation}
In order to reduce the dimension of the Hamiltonian matrix we
introduce a cutoff energy $E_{\text{cut}}$, set the maximum number of plane waves $ P=\smash{\sqrt{2 E_{\text{cut}}}/\Delta k} $ and choose the maximum number of orbital modes $N$ in Eq.~\eqref{eq:wavefunc} as the nearest integer to $n(p)= R \smash{\sqrt{2E_{\text{cut}} -(p \Delta k)^2}}$. The Hamiltonian matrix is block diagonalized using a symmetrized basis set. In particular, we consider linear combinations of the above basis functions that are even or odd with respect to the inversion of the relative coordinates $x$ and $y$, which is the equivalent of inverting the absolute coordinates, since the corresponding inversion operators $\Pi_x, \Pi_y$ commute with the relative motion
Hamiltonian
\begin{eqnarray}\label{eq:inversion_commute}
[H, \Pi_x] = 0 &,& [H, \Pi_y] = 0.
\end{eqnarray}
The resulting energy $ E_j $ is obtained with respect to the energy minimum of the conduction band. Therefore the binding energy of the exciton ground state is
$ \exb=-E_0 $.
In the presence of a photon gauge field the electron-hole pair recombines emitting a photon of energy $ E_g-\exb $. The recombination rate is related to the dimensionless oscillator strength $ f $, which in the dipole approximation reads\cite{andreani:95}
\begin{equation}\label{eq:oscill_strength}
f= S_0  \frac{ |\psi_0(0,0)|^2}{E_g-\exb}\delta_{\bm{Q},\bm{0}}.
\end{equation}
Here $ |\psi_0(0,0)|^2 $ is the envelope function of the exciton ground state given by Eq.~\eqref{eq:wavefunc}, $ \bm{Q} $ the momentum of the center of mass, $ E_g $ is the energy gap between valence and conduction band and $S_0= E_p / \epsilon_S $, where
$ E_p $ is the energy associated with Kane's matrix elements.\cite{bastard:88}
\section{Results}\label{sec:results}

In the following we investigate the excitonic properties of QTs made of the direct gap materials, InAs, GaAs and InP, and two different dielectric configurations: \emph{filled} QTs, with a core
of the same material of the shell ($\epsilon_C = \epsilon_S \neq \epsilon_E$), and
\emph{hollow} QTs, with the core of the same material as the environment ($\epsilon_C = \epsilon_E \neq \epsilon_S$).
We consider QTs with diameters in the $20\div 100$~nm range and a constant shell
thickness $ 2\delta = 10$~nm, comparable to state-of-the-art samples.\cite{lauhon:02, deneke:04, mohan:06b} Material parameters used in the calculations are listed in Table~\ref{tab:parameters}.

\begin{figure*}
\begin{center}
 \begin{tabular}{c c c}
\includegraphics[width=0.65\columnwidth,clip]{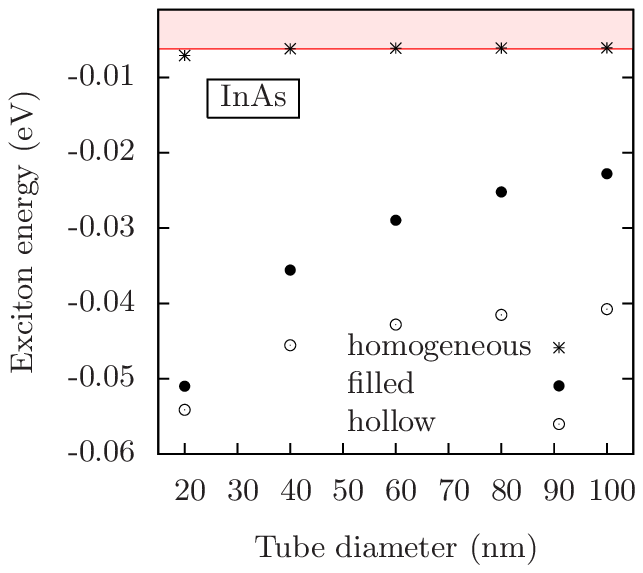} &
 \includegraphics[width=0.65\columnwidth,clip]{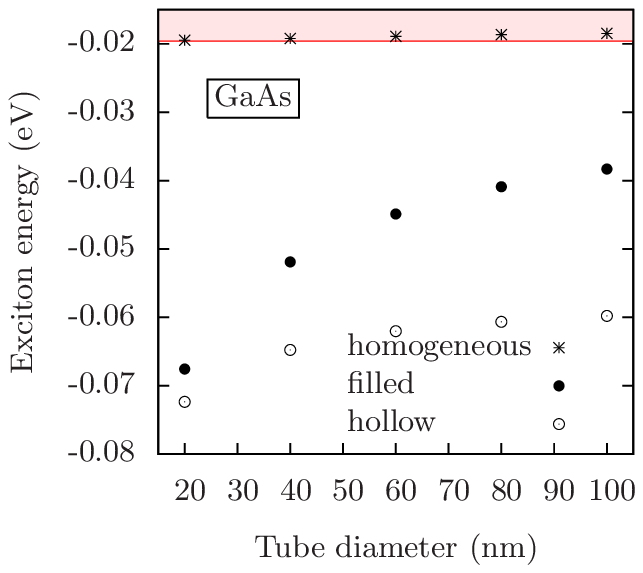} &
\includegraphics[width=0.65\columnwidth,clip]{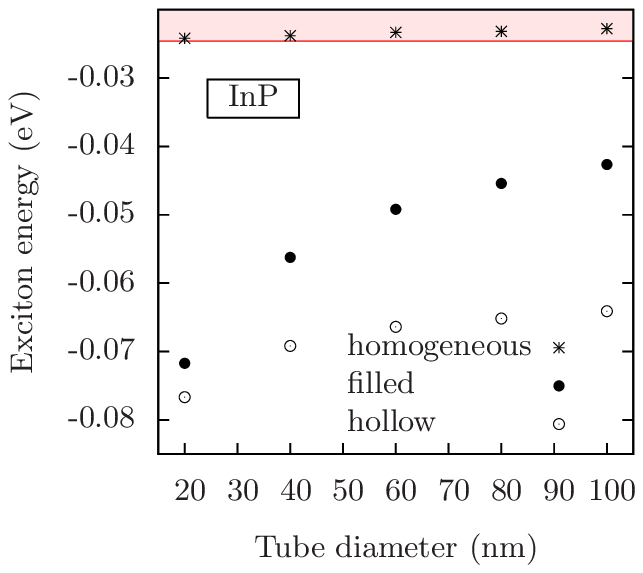}
 \end{tabular}
\caption{Ground state energy of the exciton for InAs, GaAs and InP tubes. Different
symbols correspond to a homogeneous dielectric constant ($\epsilon_C = \epsilon_S =
\epsilon_E$, crosses), a filled tube ($\epsilon_C = \epsilon_S, \epsilon_E = 1$, filled
circles) and a hollow tube ($\epsilon_C=\epsilon_E=1$, empty circles). The shaded region marks the analytical 2D limit for the homogenous case. The exciton energy for
hollow and filled QTs increases as $\sim -10/D$ and $\sim -10/D^2$, respectively.}
\label{fig:ene_vs_diameter}
\end{center}
\end{figure*}

In Fig.~\ref{fig:ene_vs_diameter} we plot the energy of the exciton for hollow
and filled QTs in vacuum ($\epsilon_E =1$), and compare it with the 2D limit
$R\rightarrow\infty$; this, in the case of excitons confined to a strictly 2D layer,
is four times the bulk value.\cite{haug:04, bastard:82, nota-2D}
The exciton binding energy shows a strong increase with respect to the 2D limit, and a marked diameter dependence, which is different for hollow and filled QTs: while in the former case the exciton binding energy becomes weakly dependent of the diameter for QTs larger
than $\sim 60$ nm, the latter one show a strong dependence even for the larger QTs.

It is instructive to contrast these results with the exciton energies of QTs which are dielectrically homogeneous, that is, buried in a material with the same bulk dielectric constant of the semiconductor shell ($\epsilon_S=\epsilon_C=\epsilon_E$). In this case, exciton energies do not show any dependence on the diameter, and are pinned to the 2D value (see Fig.~\ref{fig:ene_vs_diameter}) . This is due to the small value of the Bohr radius with respect to the tube diameters shown here, so that the curvature of the surface plays only a minor role.\cite{nota-conv} Clearly, for smaller diameters (not shown here), the exciton binding energy increases and the exciton energy red-shifts, since the binding energy is infinite in the strictly 1D limit.\cite{pedersen:03, kammerlander:07} In Fig.~\ref{fig:ene_vs_diameter} this can only be recognized in the tiny red-shift for InAs at the smallest diameter.

The large binding energy of dielectrically modulated QTs with respect to homogeneous ones is an obvious consequence of the smaller screening of the electron-hole interaction in regions where the dielectric constant is 1. Accordingly, exciton energies of filled and hollow QTs are similar for the smaller diameters, because their dielectric configuration differs only in the core, which is a small fraction of the volume if $2 \delta=10$~nm. Increasing the diameter while leaving the shell thickness constant corresponds to increasing the core region with respect to the shell, thus enhancing the difference between filled and hollow QTs. In both cases energies are increasing with diameter due to the larger area occupied by the cross-section of semiconductor shell, but faster for filled than for hollow QTs. This is consistent with the following argument: the cross section of the tube with $\epsilon_S$ is a ring of area $2D \delta \pi$ for hollow QTs and a circle of area $(D/2+\delta)^2\pi$ for filled ones. Thus, the screening area is growing faster for filled QTs. In fact, the exciton energy for
hollow and filled QTs increases as $\sim -10/D$ and $\sim -10/D^2$, respectively.

All in all, excitons in QTs can be from twice (filled GaAs QTs, $D=100$~nm) up to 7 times (hollow InAs QTs at $D=20$~nm) more strongly bound with respect to the respective 2D bulk exciton. Moreover, in all investigated cases they have a binding energy which is much larger than thermal energy at room temperature.

 \begin{figure*}
 \begin{tabular}{c c c}
  \includegraphics[width=0.65\columnwidth,clip]{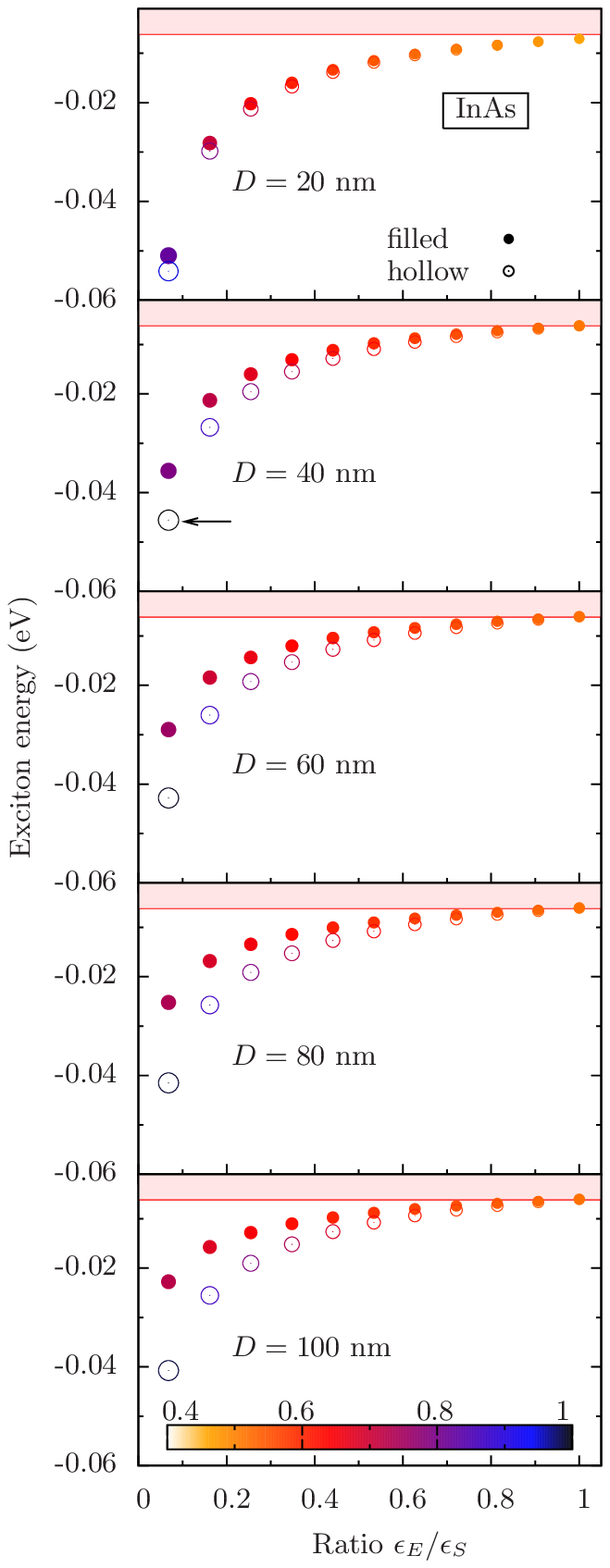} &
  \includegraphics[width=0.65\columnwidth,clip]{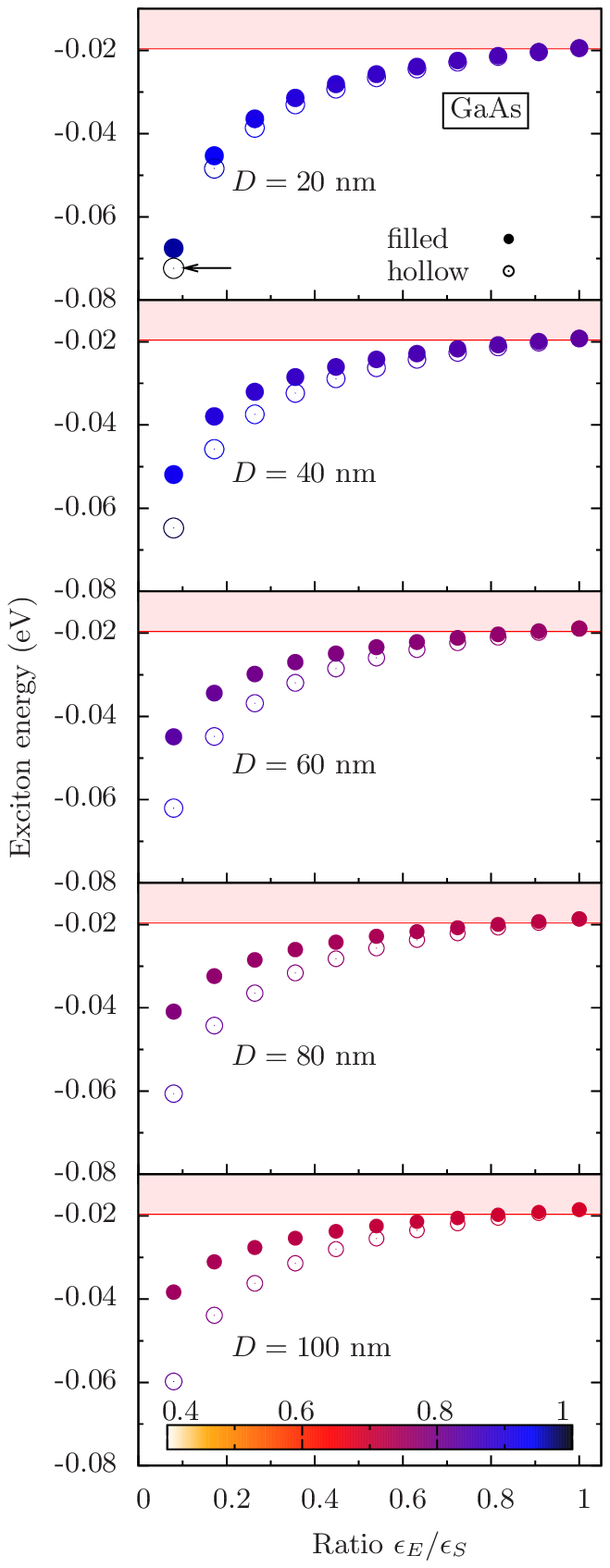} &
  \includegraphics[width=0.65\columnwidth,clip]{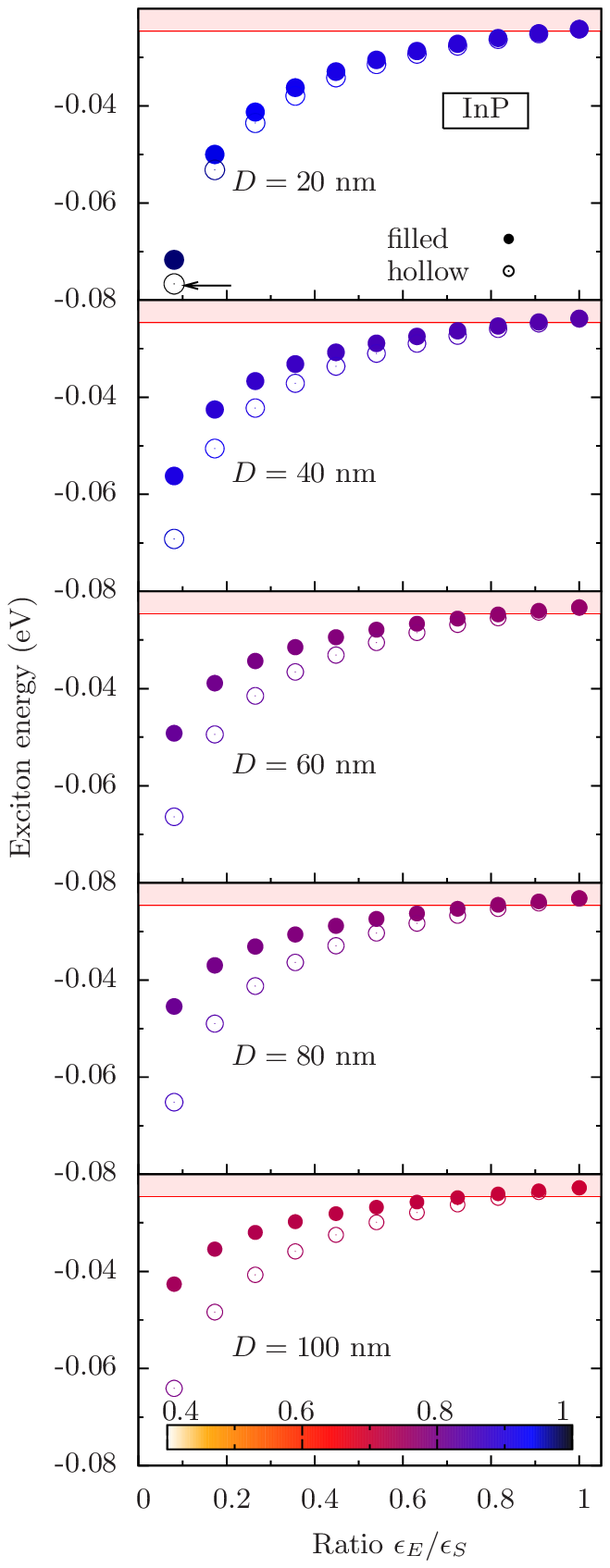}
 \end{tabular}
\caption{(color online) Ground state energy of the exciton for different values of the diameter $D$ as a function
of the dielectric contrast $ \epsilon_E / \epsilon_S$, with $\epsilon_C=\epsilon_E$ ($\epsilon_C=\epsilon_S$) for hollow (filled) QTs. The shaded area indicates the limit of excitons in a 2D quantum well. Color  and size of the symbols (filled and empty circles for the respective QTs) are proportional to the relative oscillator strength, normalized to the maximum value for each material (marked with an arrow).}
\label{fig:ene_vs_epsilon}
 \end{figure*}

In Fig.~\ref{fig:ene_vs_diameter} we used the permittivity of vacuum, which leads to the largest possible dielectric confinement effects in a given configuration. Next we show how the binding energy depends on the dielectric constant of the medium by which the QTs are surrounded.\cite{kuriyama:98, deneke:04, issac:05} Figure~\ref{fig:ene_vs_epsilon} shows the binding energy as a function of the ratio $\epsilon_E/\epsilon_S$. The leftmost point on the horizontal axis corresponds to the case $\epsilon_E =1$, while the ratio $\epsilon_E/\epsilon_S=1$ correspond to the homogeneous case, with the same dielectric constant $\epsilon_S$ filling all space. The general behavior is the same for the three materials. The difference
in energies between hollow and filled tubes is largest in vacuum for large diameters. Increasing the screening of the surrounding leads to less strongly bound excitons, since the Coulomb interaction is more and more inhibited, and the two cases of hollow and filled tubes are becoming increasingly similar to each other, and obviously coincide at $\epsilon_E/\epsilon_S =1$. We also note that the exciton binding energy is very sensitive to the environment (and core) in the low-dielectric-constant range. The exciton binding energy with respect to the 2D case is halved increasing $\epsilon_E  $ from $0.1 \epsilon_S$ to $\sim 0.2 \epsilon_S$, which suggests that small changes of the dielectric environment might be revealed by optical means in this type of system, which is due to the proximity of the electronic system to the environment.
We stress that excitons in InAs tubes are always less bound than excitons in GaAs or InP tubes for any considered value of the ratio $\epsilon_E/\epsilon_S$.

In Fig.~\ref{fig:ene_vs_epsilon} we also show in color-scale (color online) and point-size the oscillator strength of Eq.~\eqref{eq:oscill_strength} as a function of diameter and dielectric configuration. For each material the oscillator strength is normalized relative to the maximum value for the same material, so that the atomic part of the oscillator strength cancels out. The results shown in Fig.~\ref{fig:ene_vs_epsilon} can be summarized as follows: i) For GaAs and InP QTs the recombination probability is larger the smaller is the diameter, while for InAs QTs it is nearly insensitive to it; ii) for GaAs and InP QTs the relative oscillator strength is very weakly dependent on $ \epsilon_E $, while for InAs QTs it  decreases for increasing ratio $ \epsilon_E/\epsilon_S$. The peculiar behaviour of InAs with respect to GaAs and InP in both respects must be traced to interplay between the small band gap of InAs and the very large exciton binding energy in this class of systems, making the denominator in Eq.~\eqref{eq:oscill_strength} strongly dependent on $ \exb $.
\begin{table}
\caption{Parameters of materials under consideration. The mass, energy and length are in units of bare electron mass, eV and nm respectively. Values are taken from Refs.~\onlinecite{slachmuylders:07, bastard:88, ioffe}.}
\label{tab:parameters}
\begin{ruledtabular}
\begin{center}
\begin{tabular}{c c c c c c c}
 Material &$ m_e $&$ m_h $& $\epsilon_S $ & $\Ha$ & $a_B^*$  &$ E_g$  \\
\hline InAs & 0.026& 0.33 &14.6 & 0.0031 & 32.05 & 0.35 \\
 GaAs & 0.067& 0.35 &12.5 & 0.0098 & 11.76 & 1.43\\
 InP & 0.08& 0.33& 12.4 & 0.0123 & 9.23 & 1.34\\
\end{tabular}
\end{center}
\end{ruledtabular}
\end{table}

\begin{figure*}
 \begin{tabular}{c c c}
  \includegraphics[width=0.65\columnwidth,clip]{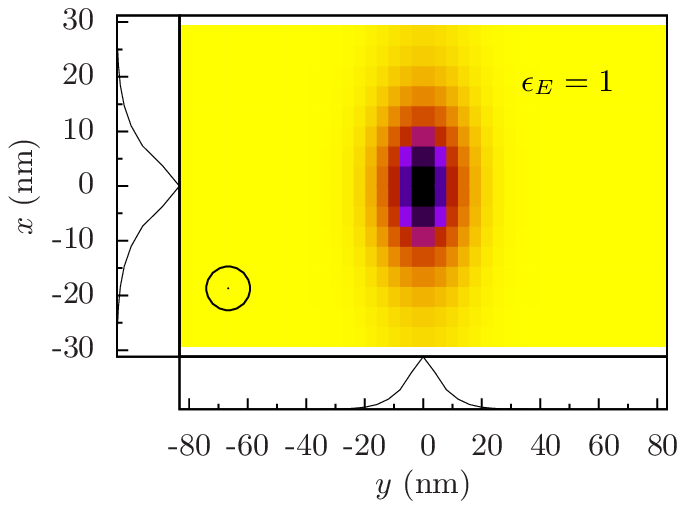}&  \includegraphics[width=0.65\columnwidth,clip]{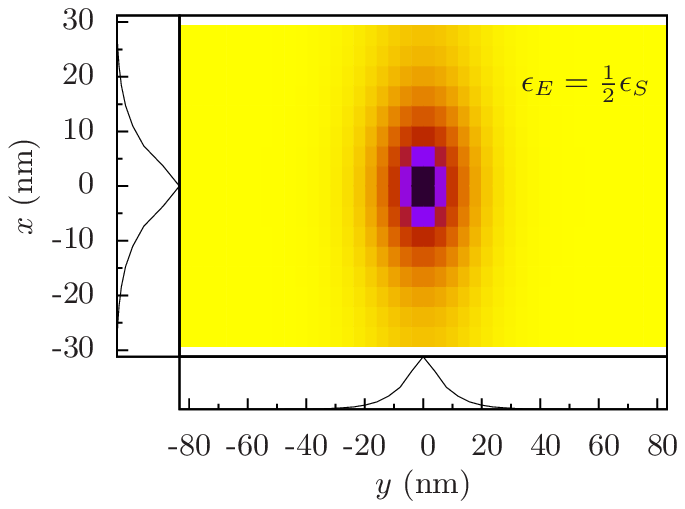}  & \includegraphics[width=0.65\columnwidth,clip]{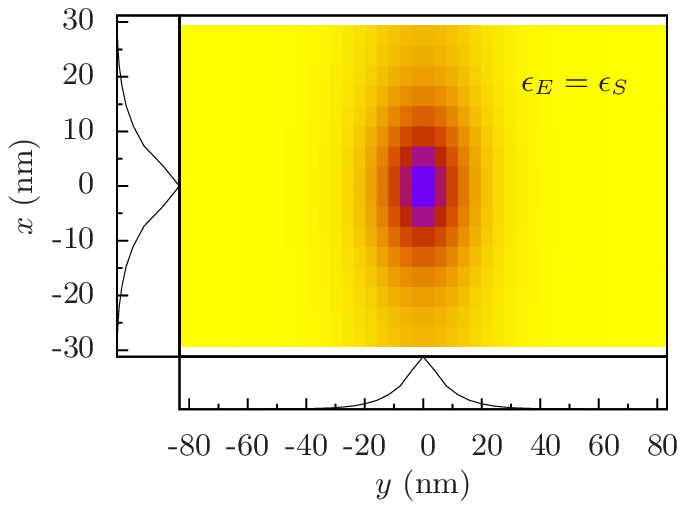} \\
     \includegraphics[width=0.65\columnwidth,clip]{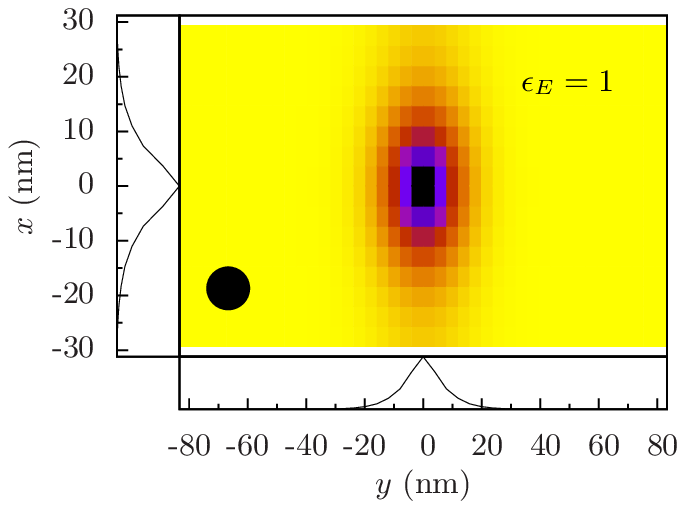}& \includegraphics[width=0.65\columnwidth,clip]{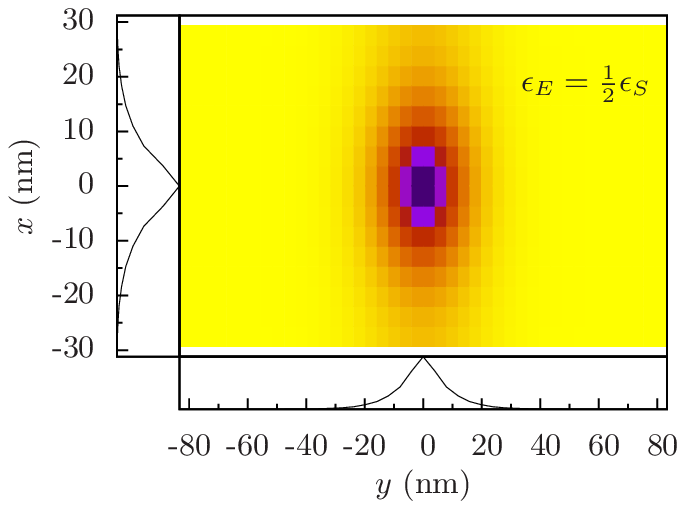} & \includegraphics[width=0.65\columnwidth,clip]{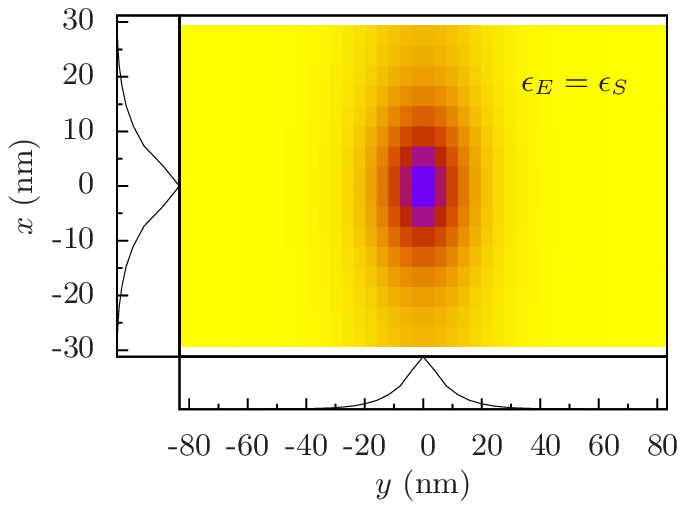}  \\
 \end{tabular}
\caption{(color online) Wave function of the exciton ground state in InAs for tubes of diameter $D=20$~nm and three different values of $ \epsilon_E $. Top row (empty circle): hollow QTs, $\epsilon_C=\epsilon_E$. Bottom row (filled circle): filled QTs, $\epsilon_C=\epsilon_S$. For each panel, an inset on the left (bottom) shows the cut of the total wave function at $y=0$ ($x=0)$. The $x$-axis extends from $-R\pi$ to $R\pi$.}
\label{fig:wave_20nm}
\end{figure*}

\begin{figure*}
 \begin{tabular}{c c c}
   \includegraphics[width=0.65\columnwidth,clip]{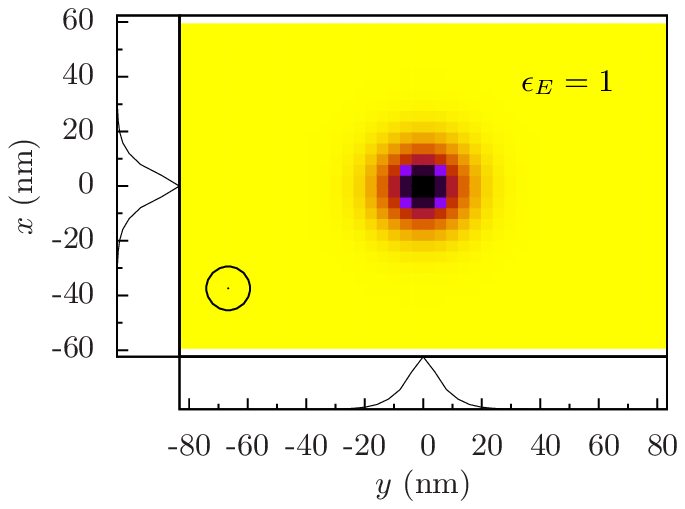}&  \includegraphics[width=0.65\columnwidth,clip]{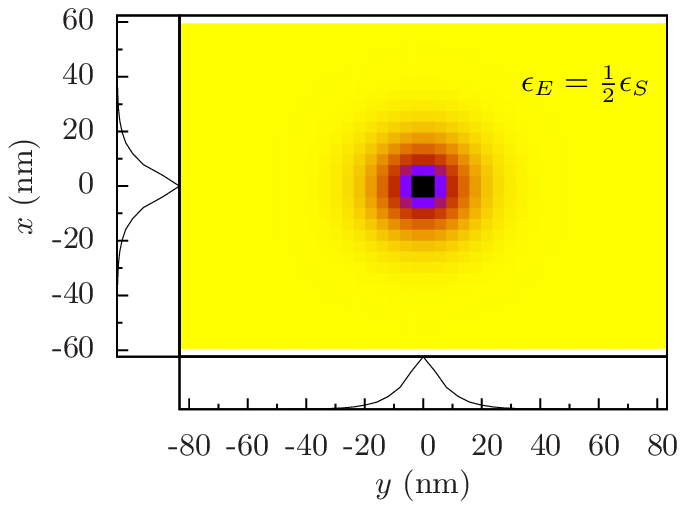}  & \includegraphics[width=0.65\columnwidth,clip]{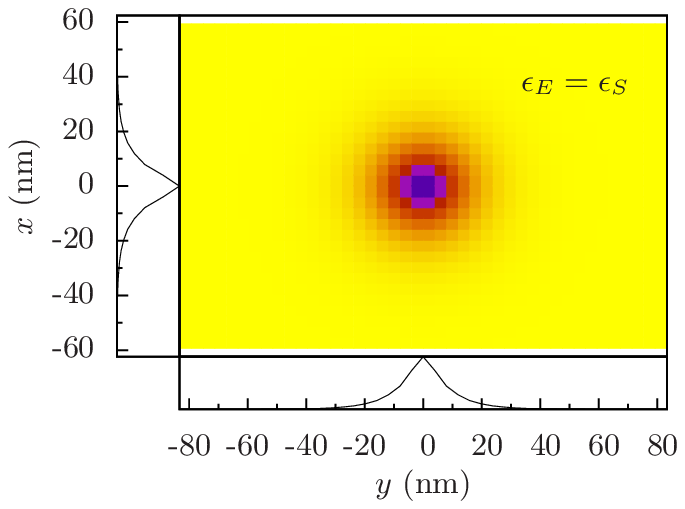} \\
     \includegraphics[width=0.65\columnwidth,clip]{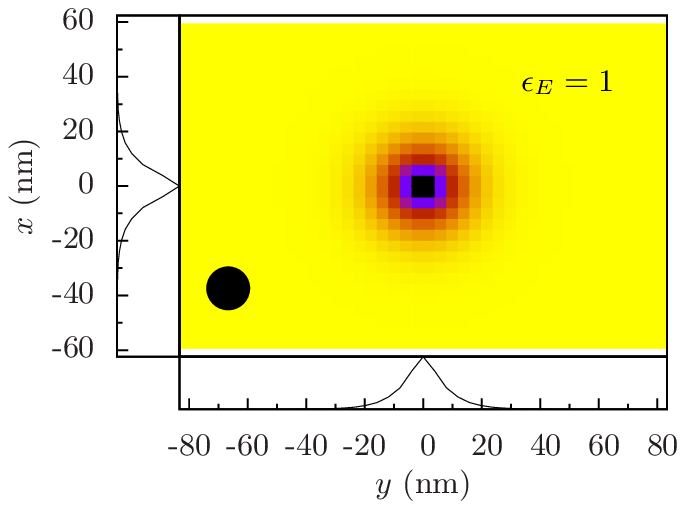}& \includegraphics[width=0.65\columnwidth,clip]{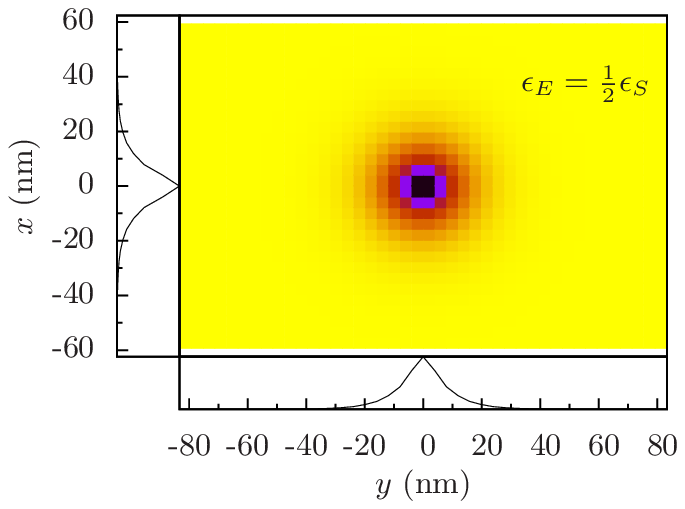} & \includegraphics[width=0.65\columnwidth,clip]{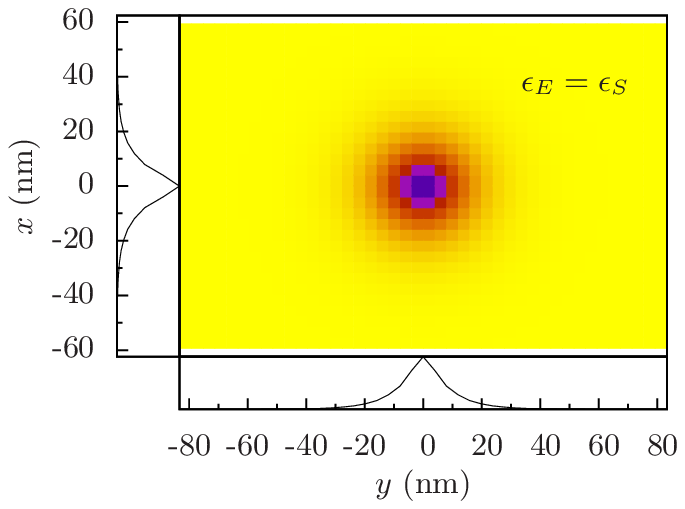}  \\
 \end{tabular}
\caption{As in Fig.~\ref{fig:wave_20nm} for tubes of diameter $D=100$~nm.}
\label{fig:wave_100nm}
 \end{figure*}

In order to further investigate the effect of the Coulomb interaction between the
carriers, we plot the squared modulus of the ground state excitonic wave function for three different dielectric environments, for a InAs tube of diameters $D=20$~nm (Fig.~\ref{fig:wave_20nm}) and $D=100$~nm (Fig.~\ref{fig:wave_100nm}), for both hollow (top) and filled (bottom) tubes. These are two relevant cases, since the former exhibits a geometrical confinement caused by the small circumference, whereas the latter falls fully in the 2D regime without any confinement.

As shown in Fig.~\ref{fig:wave_20nm}, the wave function for small tubes is distributed over all the circumference, best visible in the homogeneous case ($ \epsilon_E=\epsilon_S $). Reducing the screening by diminishing $ \epsilon_E $ affects the wavefunctions only weakly, leading to a slightly increased localization, both of the hollow (upper panels) as well as of the filled case (lower panels). On the other hand, both dielectric configurations lead to similar wavefunctions, reflected in the energies reported in Fig.~\ref{fig:ene_vs_epsilon}, too.

For large tubes of Fig.~\ref{fig:wave_100nm} the wavefunction is no more distributed over all the circumference, but well localized. Therefore the curvature of the tube has no effects on the exciton for larger diameters, making it fully 2D. Again, changing the dielectric configuration,  by diminishing $ \epsilon_E $ as well as by going from hollow (top panels) to filled (bottom panels) tubes, is changing the wavefunctions only marginally, while the respective energies are very sensitive to it (see Fig.~\ref{fig:ene_vs_epsilon}).

Therefore, while the diameter has a definite influence on the dimensionality of the excitonic states, changing the dielectric configuration amounts to modulating the mean screening with nearly no effects on the wavefunction, shifting only the energy.

\section{Conclusions}

We have studied theoretically the excitonic properties of semiconductor QTs, focusing on the influence of their dielectric environment and its interplay with structural parameters. We find that, due to the strong increase of the electron-hole interaction and ensuing very large excitonic binding which is possible in these structures, the spectral properties of excitonic absorption are strongly dependent on geometrical parameters and dielectric environment, with energies well below the energies of the dielectrically homogenous case which is always in the 2D regime for typical parameters. Calculations have been performed for InAs, GaAs and InP. The low gap material InAs shows a peculiar behavior, since in the investigated systems the exciton binding energy is a substantial fraction of the gap. The very large binding energies, their tunability in a wide range, and the large sensitivity of the excitonic response to the dielectric medium, point to perspective applications of these systems.

\section{Acknowledgments}

We thank financial support from the Italian Minister for University and Research through FIRB RBIN04EY74 and CINECA Iniziativa Calcolo Parallelo 2009.

\appendix*
\section{Derivation of the Coulomb interaction in a tube}\label{sec:potential}
The inner radius $a=R-\delta$ and the outer radius $b=R+\delta$ divide the space into three regions: core ($\rho <a$), shell ($a<\rho<b$) and environment ($\rho> b$) with dielectric constants $\epsilon_C$,  $\epsilon_S$, $\epsilon_E$, respectively (see Fig. \ref{fig:tube_scheme}).
The electrostatic potential at point $\bm{r}$ induced by an electron localized in the shell, \textit{i.e.} with $ a\le \rho' \le b $, screened by $ \epsilon_S $ has to obey the Poisson equation in cylindrical coordinates (with charge $e=1)$:
\begin{align}\label{eq:poisson}
\nabla^2_{\bm{r}}V_{\alpha}(\bm{r},\bm{r'})&= -\frac{4 \pi}{\epsilon_S \rho}\delta(\rho-\rho') \delta(\phi-\phi') \delta(z-z').
\end{align}
Here $ {\alpha}$ indicates one of the three possible regions of the test charge: core ($C$), shell ($ S $) or environment ($ E $).
Eq. \eqref{eq:poisson} is solved by the \emph{ansatz}
\begin{align}\label{eq:green_func_real}
V_{\alpha}(\bm{r},\bm{r'})&= \frac{1}{2 \pi^2} \sum_{m=-\infty}^{\infty} e^{\imath m (\phi-\phi')} \times \nonumber\\
&\int_0^{\infty}dk \cos(k(z-z'))g_{m,{\alpha}}(k, \rho, \rho'),
\end{align}
where $g_{m,{\alpha}}(k, \rho, \rho')$ is the solution of the radial Poisson equation in each region ${\alpha}$
\begin{equation}\label{eq:radial_poisson}
\frac{1}{\rho}\frac{\partial}{\partial \rho}\bigg (\rho \frac{\partial g_{m,{\alpha}}}{\partial \rho} \bigg)-\bigg (k^2+\frac{m^2}{\rho^2}\bigg )g_{m,{\alpha}}=-\frac{4\pi}{\epsilon_S \rho}\delta(\rho-\rho'),
\end{equation}
and can be written as a linear combination of the solutions of the homogeneous Laplace equation, \textit{i.e.} modified Bessel functions of the first kind, $I_m(q \rho)$, and the second kind, $K_m(q \rho)$, with the following properties:\cite{abramowitz:72}
\begin{subequations}
\begin{eqnarray}
\lim_{x \rightarrow 0}&I_m(x)= 0, & \lim_{x \rightarrow \infty}I_m(x)= \infty, \\
\lim_{x \rightarrow 0}&K_m(x)= \infty,  & \lim_{x \rightarrow \infty}K_m(x)= 0.
\end{eqnarray}
\end{subequations}
Imposing that $ \lim_{\rho \rightarrow \infty} g_{m,E}(k \rho)= 0$ and the finiteness of $ \lim_{\rho \rightarrow 0} g_{m,C}(k \rho)$, we have
\begin{subequations}\label{eq:green_fu_3_region}
\begin{align}
g_{m,C}(\rho)&= A_m I_m(k\rho)  \\
\gamma_{m,S}^{</>}(\rho^{</>})&=B_m^{</>}I_m(k\rho^{</>})+C_m^{</>}K_m(k\rho^{</>})  \\
g_{m,E}(\rho)&= D_m K_m(k \rho),
\end{align}
\end{subequations}
where $ \gamma_{m,S}^{</>} $ are no Green's functions, but solutions of the (homogeneous) Laplace equation, from which we construct the solution $g_{m,S}(\rho, \rho')$ of Eq.~\eqref{eq:radial_poisson} in the following. We define $\rho^< = \text{min}[\rho, \rho']$ and $\rho^> = \text{max}[\rho, \rho']$. 
Matching components of fields $E_{\parallel} $ and $D_{\perp}$ at the interfaces is equivalent to~\cite{abajo:02} 
\begin{subequations}\label{eq:bound_condition}
\begin{align}
g_{m,C} & = \gamma_{m,S}^{<}, 
& \epsilon_C \frac{\partial g_{m,C}}{\partial \rho} & =  \epsilon_S \frac{\partial \gamma_{m,S}^<}{\partial \rho} 
& \text{at       } \rho=a, \\
g_{m,E} & = \gamma_{m,S}^{>} , 
&   \epsilon_E \frac{\partial g_{m,E}}{\partial \rho}  & = \epsilon_S \frac{\partial \gamma_{m,S}^>}{\partial \rho}
& \text{at       } \rho=b.
\end{align}
\end{subequations}
To determine the last two unknowns we use the symmetry of the Green's function $g_{m,S}(\rho, \rho')$ with respect to the exchange of  $\rho$ and $\rho'$ making,\cite{jackson:98}
\begin{equation}\label{eq:green_fu_region_2}
g_{m,S}(\rho, \rho')=\gamma_{m,S}^< \gamma_{m,S}^>,
\end{equation}
and normalization
\begin{equation}\label{eq:green_fu_norm}
\gamma _{m,S}^<(\rho)\frac{d \gamma_{m,S}^>(\rho)}{d\rho} 
-\gamma_{m,S}^>(\rho) \frac{d \gamma_{m,S}^<(\rho)}{d \rho}=-\frac{4 \pi}{\epsilon_S \rho }.
\end{equation}
Defining $ \alpha= \epsilon_S / \epsilon_C $ and $ \beta = \epsilon_E / \epsilon_S $, and the quantities
\begin{subequations}
\begin{align}
S_m^l(k\rho) & = \frac{K_m(k\rho)I_m(k\rho)'}{I_m(k\rho) K_m(k\rho)'}\bigg \vert _{\rho=l}, \\
T_m^l(k\rho) & = \frac{K_m(k\rho)}{I_m(k\rho)}\bigg \vert_{\rho=l}, \\
U_m & = T_m^a (S_m^a-\alpha)(S_m^b - \beta) - S_m^a T_m^b (\alpha-1)(\beta-1),
\end{align}
\end{subequations}
where $I(x)'= dI(x)/ dx$ and $l={a,b}$ indicates the inner and outer radius of the cylindrical shell,
the coefficients in Eq.~\eqref{eq:green_fu_3_region} are given by
\begin{subequations}\label{eq:green_fu_coeff}
\begin{align}
A_m&=\frac{4 \pi}{\epsilon_S} \alpha T_m^a (S_m^a-1)(S_m^b-1)/U_m, \\
B_m^<&= \frac{4 \pi}{\epsilon_S} T_m^a (S_m^b-1)(S_m^a- \alpha)/U_m, \\
B_m^>&= T_m^b\frac{\beta-1}{S_m^b-1}, \\
C_m^<&= \frac{4 \pi}{\epsilon_S} S_m^a (S_m^b-1)(\alpha-1)/U_m, \\
C_m^>&= \frac{S_m^b-\beta}{S_m^b-1}, \\
D_m&=1
\end{align}
\end{subequations}

\noindent 
In particular, for two charges localized at the same distance $\rho=\rho^\prime=R$ from the center
\begin{align}\label{eq:green_fu_shell}
g_{m,S}(k,R,R)= & \left [B_m^<I_m(kR)+ C_m^<K_m(kR) \right] \times  \nonumber\\
   & \left [B_m^> I_m(kR) + C_m^> K_m(kR) \right] \nonumber \\
   =&\frac{4 \pi}{\epsilon_S} \left[ \tilde{B}_m^< + \tilde{C}_m^<\right] \left[\tilde{B}_m^>+C_m^>\right] \times \nonumber \\
   & I_m(kR) K_m(kR),
\end{align}

\noindent where we defined for clarity

\begin{subequations}\label{eq:substit_c_b}
\begin{align}
\tilde{B}_m^< & = \frac{\epsilon_S}{4 \pi} B_m^<, \\
\tilde{C}_m^< & =\frac{\epsilon_S}{4 \pi} \frac{K_m(kR)}{I_m(kR)} C_m^<, \\
\tilde{B}_m^> & =  \frac{I_m(kR)}{K_m(kR)}B_m^>.
\end{align}
\end{subequations}

\noindent Hence, taking Eq.~\eqref{eq:green_func_real} in the special case $\rho=\rho^\prime=R$, gives the Coulomb potential for two particles localized in the shell on a cylindrical surface of radius $R$
\begin{align}\label{eq:potential_real}
V(\bm{r}, \bm{r'})&= \frac{2}{ \pi \epsilon_S} \sum_{m=-\infty}^{\infty} e^{\imath m (\phi-\phi')} \times \nonumber \\
   &\int_0^{\infty} \left[ \tilde{B}_m^< + \tilde{C}_m^<\right] \left[\tilde{B}_m^>+C_m^>\right]  \times \nonumber \\ &I_m(kR) K_m(kR) \cos(k(z-z')) dk.
\end{align}
For $\epsilon_C=\epsilon_S =\epsilon_E$ this reduces to the usual form $1/\epsilon_S |r-r^\prime|$ in cylindrical coordinates,\cite{jackson:98} while for $\epsilon_C=\epsilon_S \neq \epsilon_E$ Eq.~\eqref{eq:potential_real} reproduces the result of Ref.~\onlinecite{byczuk:99}. Note that $V$ is scalable, since all arguments in Eq. \eqref{eq:potential_real} are products of lengths and momenta and thus dimensionless, only the measure $dk$ of the integral is reciprocal in length. The latter one scales with the effective bohr length a$_B^*=(\epsilon_S/\mu) 0.053$~nm and therefore $V$ itself with the effective Hartree \Ha$=(\mu/\epsilon_S^2) 27.21$~eV.


\begin{thebibliography}{45}
\bibitem[{not({\natexlab{b}})}]{nota-2D}
\bibinfo{note}{Although for a finite confinement the 2D binding energy would be
  somewhat smaller, due to exciton spill-out in the barrier material, we expect
  that the relative difference between the QT and its 2D limit is less
  sensitive to the details of the confinement, since it is mainly due to
  dielectric effects.}
\bibitem[{not({\natexlab{c}})}]{nota-conv}
\bibinfo{note}{Small deviations from the 2D limit for GaAs and InP tubes at
  large diameters in Fig.~\ref{fig:ene_vs_diameter} are a numerical artifact due
  to less convergence in this extreme 2D limit.}
\end{thebibliography}
\end{document}